\documentclass[11pt]{article}
\usepackage{amsmath,amssymb,color,graphics,epsfig}
\usepackage{amsfonts}
\newcommand{\hoch}[1]{$\, ^{#1}$}

\newcommand{\be}{\begin{equation}}
\newcommand{\ee}{\end{equation}}
\newcommand{\bea}{\setlength\arraycolsep{2pt} \begin{eqnarray}}
\newcommand{\eea}{\end{eqnarray}}

\def\ft#1#2{{\textstyle{\frac{\scriptstyle #1}{\scriptstyle #2} } }}
\def\fft#1#2{{\frac{#1}{#2}}}

\def\0{{\sst{(0)}}}
\def\1{{\sst{(1)}}}
\def\2{{\sst{(2)}}}
\def\3{{\sst{(3)}}}
\def\4{{\sst{(4)}}}
\def\5{{\sst{(5)}}}
\def\6{{\sst{(6)}}}
\def\7{{\sst{(7)}}}
\def\8{{\sst{(8)}}}
\def\sst#1{{\scriptscriptstyle #1}}

\def\R{{\mathbb{R}}}

\begin{document}

\begin{flushright}
%\hfill{KIAS-P12028}
 %\hfill{
%\bf hep-th/yymmnnn}
\end{flushright}

\vspace{25pt}
\begin{center}
{\large {\bf Scalar Hairy Black Holes in General Dimensions}}

\vspace{10pt}
Xing-Hui Feng\hoch{1}, H. L\"u\hoch{1} and Qiang Wen\hoch{2}

\vspace{10pt}

\hoch{1} {\it Department of Physics, Beijing Normal University,
Beijing 100875, China}

\vspace{10pt}

\hoch{2} {\it Department of Physics, Renmin University of China, Beijing 100872, China}

\vspace{40pt}

\underline{ABSTRACT}
\end{center}

   We obtain a class of asymptotic flat or (A)dS hairy black holes in $D$-dimensional Einstein gravity coupled to a scalar with certain scalar potential.  For a given mass, the theory admits both the Schwarzschild-Tangherlini and the hairy black holes with different temperature and entropy, but satisfying the same first law of thermodynamics.  For some appropriate choice of parameters, the scalar potential can be expressed in terms of a super-potential and it can arise in gauged supergravities.  In this case, the solutions develop a naked curvature singularity and become the spherical domain walls.  Uplifting the solutions to $D=11$ or 10, we obtain solutions that can be viewed as spherical M-branes or D3-branes. We also add electric charges to these hairy black holes.  All these solutions contain no scalar charges in that the first law of thermodynamics are unmodified.  We also try to construct new AdS black holes carrying scalar charges, with some moderate success in that the charges are pre-fixed in the theory instead of being some continuous integration constants.

\thispagestyle{empty}

\pagebreak
%\voffset=0pt
%\setcounter{page}{1}

%\tableofcontents
%\addtocontents{toc}{\protect\setcounter{tocdepth}{2}}

%%%%%%%%%%%%%%%%%%%%%%%%%%%%%%%%%%%%%%%%

\newpage
%%%%%%%%%%%%%%%%%%%%%%%%%%%%%%%%%%%%%%%%

\section{Introduction}

Since the discovery of Einstein's General Relativity, there have been continuing
efforts in constructing new exact solutions.  The black holes are of particular
interest since they possess many intriguing properties such as the Hawking
radiation, holography on the horizon, thermodynamics due to the semi-classical
effect and the no-hair phenomenon {\it etc.}  With the advance of string theory
and its low-energy effective supergravities, large number of exact black hole
solutions have been found in these theories.  The notable examples are the
multi-charge asymptotic Anti-de Sitter (AdS) black holes in the gauged
supergravity STU-like models \cite{Behrndt:1998jd,Duff:1999gh,%
tenauthor,Cvetic:1999un,Klemm:2012yg}\footnote{The STU-model is the
$SL(2,\R)^3$-invariant theory in four-dimensions, involving the $U(1)^4$
gauge fields \cite{stumodels}.  We refer to theories involving abelian
gauge fields truncated from (gauged) supergravities as STU-like models}.
Some of these solutions has been generalized to arbitrary dimensions
\cite{Wu:2011zzh,Chow:2011fh,Liu:2012ed,Lu:2013eoa}.

Few examples of exact black hole solutions are known involving only the scalar matter fields.  Although gauged supergravities may contain a large number of scalar fields with non-trivial scalar potentials, they are typically used to construct solutions of domain walls, which has the analogous geometry as that in \cite{Cvetic:1996vr}. For planar domain walls, owing to the existence of super-potentials, it is straightforward to obtain the supersymmetric solutions by solving some appropriate first-order equations.  The spherical domain walls were much more difficult to construct and few examples of exact solutions exist \cite{Chamblin:1999ya}.

The multi-charge AdS black holes of the STU-like models involve not only various vector fields, but also a set of scalar fields with some specific supergravity scalar potentials.  One would expect that in suitable limit, one could arrive at some neutral black hole solutions supported by purely the scalar fields.  For the spherical topology, it turns out that the limit of turning off the charges has also the effect of turning off the scalars, leading only to the Schwarzschild AdS black hole. For the planar topology, there exists a supersymmetric limit in which the charged black holes become the neutral domain walls \cite{tenauthor}, which turn out to be describable as distributed D3-branes or M-branes when lifted to higher dimensions \cite{Kraus:1998hv,Freedman:1999gk,Cvetic:1999xx}.

The difficulty of constructing a scalar black hole is not surprising.
Although a scalar may be the simplest field in any field theory, there are
great degrees of freedom in constructing a scalar potential without violating
any essential symmetry or introducing higher derivatives.  Thus if one starts
with a random scalar potential, the chance of finding an exact black hole
solution is almost null. Recently, there has been progress in constructing
charged scalar hairy black holes in four dimensions
\cite{Anabalon:2012ta,Anabalon:2013qua,Anabalon:2013sra,Gonzalez:2013aca}. In these works, instead of considering a specific scalar potential, the solution of the scalar field was assumed. Making use of some clever ansatz, one could derive all the metric functions and the scalar potential from the equations of motion. This allows one to obtain a class of exact hairy black holes with or without electric or magnetic charges.  Some hairy planar black holes in higher dimensions were also constructed in \cite{Acena:2013jya}.

The purpose of this paper is to generalize these results in general dimensions, but with a different motivation and in a different coordinate system.  In section 2, we motivate our ansatz by the usual construction of black $p$-branes including the AdS black holes in the STU-like models.  Static and isotropic black $p$-branes with regular horizon involve typically two functions which we shall call $H$ and $f$, in addition to the explicit radial $r$ dependence.  It was well-known that for the black holes in the STU-like models with or without the scalar potential, there is a direct relation between the scalar fields and the metric function $H$, which is independent of $f$ and the scalar potential.  In fact the scalar potential enters the solution only through the function $f$.  This motivates us to write down an ansatz such that given either $H$ or the scalar $\phi$, one can derive the remaining fucntions and the scalar potential in a consecutive manner.  The advantage of our ansatz is that it fits already the structure of the standard black $p$-branes and hence the asymptotic infinity is simply at $r=\infty$.

In section 3, we use a simple ansatz for the scalar $\phi$ inspired by $p$-branes, and derive the full local solutions and obtain the relevant scalar potentials.  The general solutions involve some hypergeometric functions in general dimensions, which become polynomials in four and five dimensions.  We study the scalar potentials and find that with suitable choice of parameters, it recovers the general results obtained in \cite{Lu:2013eoa}.  These potentials can be expressed in terms of super-potentials and can arise from gauged supergravities in low-lying dimensions.  We perform the global analysis of the local solutions and show that hairy black holes indeed exist. We derive the first law of thermodynamics.  Although we concentrate our attention on the spherical solutions, we also give the results for the torus and hyperbolic topologies.

For the scalar potentials that can be expressed in terms of the super-potentials, we find that the solutions contain naked curvature singularity.  The solutions with the planar geometry was known previously as (planar) domain walls .  Our results provide first many examples of spherical domain walls in gauged supergravities.  Lifting these solution to $D=11$ and $D=10$, they become hard-sought spherical M-branes and D3-branes. We discuss this in section 4.

In section 5, we add two vectors fields with some exponential dilaton couplings and obtain hairy charged solutions.  They include all the solutions obtained in \cite{Lu:2013eoa} that contain the charged AdS black holes in STU-like models.

In section 6, we show that all these hairy black holes do not have scalar charge.  This can be seen already that the first law of thermodynamics does not contain any contribution from the scalars.  Motivated by the criteria obtained in \cite{Lu:2013ura}, we try to construct hairy black holes with scalar charges, but with only some moderate success. We conclude our paper in section 7.

\section{The set up}

In a typical construction of isotropic black $p$-branes, the metric ansatz is given by \cite{Duff:1996hp,Cvetic:1996gq}
\be
ds^2=e^{2A} (-f dt^2 + dx^i dx^i) + e^{2B} (f^{-1} dr^2 + r^2 d\Omega_{D-d-1}^2)\,,
\ee
where $d=p+1$ and the metric functions $A$, $B$, $f$  and the involved scalars $\phi_i$ are functions of $r$ only. In order for the solution to have a regular horizon, one needs to impose a relation
\be
(p+1) A + (D-p-3)B=0\label{ABrelation}
\ee
that is consistent with the equations of motion. The vanishing of $f$ at certain $r=r_0$ gives rise to the horizon whilst the functions $A$ and $B$ run smoothly from the horizon to the asymptotic infinity.  It turns out the dilaton scalars $\phi_i$ depend on the function $A$ but it is independent of the function $f$.  In the case of charged AdS black holes in gauged supergravities, the contribution from the scalar potentials only modifies the function $f$, but leaves the relation between $A$ and $\phi_i$ unchanged.  This suggests that the relation between $A$ and $\phi_i$ can be determined independent of a scalar potential.  This observation allows us to construct a class of scalar hairy black holes.

    To make this concrete, let us consider Einstein gravity in general
dimensions coupled to a scalar $\phi$, with a scalar potential:
\be
e^{-1} {\cal L} = R - \ft12 (\partial\phi)^2 - V(\phi)\,,
\ee
where $e=\sqrt{-\det(g_{ij})}$.  The equations of motion are
\be
\Box\phi = \fft{dV}{d\phi}\,,\qquad Z_{\mu\nu}\equiv R_{\mu\nu} - \ft12\partial_\mu\phi\partial_\nu\phi -\ft1{D-2} V g_{\mu\nu}=0\,.\label{purescalareom}
\ee
Inspired by the black $p$-brane ansatz with the constraint (\ref{ABrelation}), we consider the following static and spherically symmetric black hole ansatz ($p=0$):
\be
ds^2=-H^{-1} f dt^2 + H^{\fft{1}{D-3}} \Big(\fft{dr^2}{f} + r^2d\Omega_{D-2}^2\Big)\,.
\ee
where $H=H(r)$ and $f=f(r)$.  It should be emphasized that this is the most general ansatz for the static and spherically-symmetric geometries for $p=0$.
We find that the specific combination $Z^t{}_t - Z^r{}_r=0$ of the Einstein equations of motion implies
\be
\fft{2(D-3)}{D-2}\phi'^2 =\fft{H'^2}{H^2} - \fft{2H''}{H} - \fft{2(D-2)H'}{rH}\,.\label{eom1}
\ee
Thus we see indeed that $H$ and $\phi$ can be determined from each other, independent of the function $f$ and the scalar potential.  We can thus start with an ansatz for $(\phi,H)$ that satisfies (\ref{eom1}).  The function $f$ can be solved from the combination $Z^t{}_t - Z^i{}_i=0$, where the index $i$ denotes any specific sphere direction. We find
\be\fft{H''}{H}+\fft{H'}{H} \Big(\fft{f'}{f}-\fft{H'}{H} + \fft{D-2}{r}\Big)- \fft{(D-3)}{(D-2)r^2f}\Big(r^2f'' + (D-4) r f' - 2(D-3)(f-1)\Big)=0\,,\label{eom2}
\ee
This is a second-order linear equation with a source for $f$ and can be solved in a variety of situations.  The scalar potential can be determined by the remaining Einstein equations, given by
\be
V=-\fft{r^2f'' + (3D-8) rf' + 2(D-3)^2(f-1)}{2r^2 H^\fft{1}{D-3}}\,.\label{eom3}
\ee
Since we are considering the most general ansatz under the static and spherical symmetry, the full set of equations of motion must involve three second-order equations for $H,f$ and $\phi$, together with a first-order Hamiltonian constraint.  This implies that the dilaton equation of motion is automatically satisfied provided that the three independent Einstein equations (\ref{eom1}), (\ref{eom2}) and (\ref{eom3}) are satisfied.  Thus provided with an ansatz for $\phi$ or $H$, the remaining functions and the scalar potential can be solved consecutively.  Note that if we make a coordinate transformation
\be
\fft{dr}{H r^2} = \eta dx\,,\label{rtox}
\ee
and define $\Omega= r^2 H^{1/(D-3)}$, we recover the ansatz proposed in
\cite{Anabalon:2012ta}.  The advantage of our ansatz is that asymptotic region of our solution is simply at $r=\infty$ and that the results can compare easily with the previously-know black holes in the STU-like models.

    In the usual construction of black holes, one starts a specific
theory and then proposes some ansatz and proceeds to solve the equations of motion.  However, the choice for a scalar potential is rather unrestricted. For a generic scalar potential, it is then very difficult if not impossible to construct exact black hole solutions.  This is the reason that there have been few exact solutions for scalar black holes in the literature.

    In this paper, instead of considering some specific choice of the
scalar potentials, we start with some reasonable simple assumption for the
solution of the scalar, as in the case of \cite{Anabalon:2012ta,Anabalon:2013qua,Anabalon:2013sra,Gonzalez:2013aca}. We can determine what $H$ must be from (\ref{eom1}) and hence the function $f$ from (\ref{eom2}).  Finally from (\ref{eom3}), we obtain the scalar potential that is responsible for the black hole.  There is however one issue about which we have to  be cautious.  The scalar potential obtained directly from (\ref{eom3}) is a function of the radial variable $r$.  We must then convert the radial dependence from $\phi(r)$ to obtain $V(\phi)$.  Thus, if we start with a specific function $\phi(r)$ involving a constant $q$, the conversion may give rise to a potential $V(\phi)$ that also depends on $q$ explicitly.  If this happens, we cannot regard the parameter $q$ as an integration constant in the solution, but rather it is a pre-fixed parameter in the Lagrangian.  Such a parameter cannot participate in the black hole first law of thermodynamics.  In our construction, we are mainly interested in solutions with parameters that do not explicitly appear in the scalar potential and hence they are true integration constants.

\section{Scalar hairy black holes}

\subsection{Local solutions in general dimensions}

In supergravities with flat spacetime vacua, a linearized scalar satisfies the massless equation $\Box\phi=0$. It falls off as
\be
\phi=\alpha + \fft{\beta}{r^{D-3}} + \cdots\,.
\ee
In gauged supergravities with AdS vacua, a linearized scalar satisfies instead
\be
(\Box +2(D-3)g^2) \phi=0\,,\label{kgeom}
\ee
where $1/g=\ell$ is the radius of the AdS.  It falls off as
\be
\phi=\fft{\alpha}{r^2} + \fft{\beta}{r^{D-3}} + \cdots\,.
\ee
In this paper, we call such a scalar massless since when embedded in (gauged) supergravities, the scalar may belong to the same super-multiplet as the graviton.

We are now in the position to construct explicit scalar hairy black holes.  To make contact with (gauged) supergravity solutions, we are interested in a scalar that is massless.  Furthermore, we require that the solution have a smooth $g=0$ limit.
The most general solution of $\phi$ at the large $r$ expansion must then take the form
\be
e^{\phi} = 1 + \fft{q}{r^{D-3}} + \cdots\,,
\ee
Inspired by the $p$-brane construction, we consider the ansatz
\be
\phi = \sqrt{\ft{D-2}{2(D-3)}}\, \nu \log H_1\qquad \hbox{with}\qquad H_1=1 + \fft{q_1}{r^{D-3}}\,,
\ee
where $\nu$ is a constant that parameterizes some degrees of choices.
It follows from (\ref{rtox}) that this implies that $\phi\sim \log x$, the ansatz used in \cite{Anabalon:2012ta,Anabalon:2013qua,
Anabalon:2013sra,Gonzalez:2013aca}. Substituting this to (\ref{eom1}), we find
\be
H=(1-c + c\, H_1^{-\mu})^2 H_1^{1+\mu}\,,\qquad \mu^2 + \nu^2=1\,.\label{h1only}
\ee
where $c$ is an integration constant.  Here we have fixed the other integration constant so that $H$ approaches 1 for $r\rightarrow \infty$. For $\mu=0$ and hence $\nu=\pm 1$, the general solution is $H= H_1 + c\, H_1\log H_1$.

     We may consider a more general possibility, namely
\be
\phi = \sqrt{\ft{D-2}{2(D-3)}}\, \nu \log \ft{H_1}{H_2}\qquad \hbox{with}\qquad H_i=1 + \fft{q_i}{r^{D-3}}\,.\label{h1h2}
\ee
The general solution for $H$ is then given by
\be
H=H_1^{1+\mu}H_2^{1+\mu} (c\, H_1^{-\mu} + (1-c)\, H_2^{-\mu})^2\,,\qquad
\mu^2 + \nu^2 = 1\,.\label{Hh1h2}
\ee
In fact, this choice of $\phi$ with both $H_1$ and $H_2$ is equivalent to that with only single $H_1$, since
\be
\fft{H_1}{H_2} = 1 + \fft{q_1-q_2}{r^{D-3} + q_2}=1 + \fft{\tilde q_1}{\tilde r^{D-3}}\,,
\ee
where $\tilde q_1=q_1-q_2$ and $\tilde r = r H_2^{1/(D-3)}$.  Using the tilded variable and coordinates, we recover the single $H_1$ case.  Thus the ansatz (\ref{h1h2}) over parameterizes; however, it has an advantage that we can turn off the scalar field $\phi$ by setting $q_1=q_2$ while retaining the non-trivial behavior for $H$.  With the single $H_1$ ansatz, if we turn off $\phi$ by turning off $q_1$, the solution becomes simply vacuous. Of course, a more subtle limit involving coordinate transformation is possible, which effectively is the same as introducing $q_2$.  For this and another reason that will become apparent in section 4, we shall proceed using (\ref{h1h2}) with both $q_1$ and $q_2$ turned on. (In fact, this phenomenon of turning on two $H_i$ being equivalent to turning on just one was already observed in some planar domain wall solutions \cite{Cvetic:1999xx}.)

     For the generic choice of $c$ in (\ref{Hh1h2}), we find that
(\ref{eom2}) cannot be solved analytically.  We shall consider a simpler choice, $c=0$, corresponding to
\be
H=H_1^{1+\mu} H_2^{1-\mu}\,.
\ee
The $c=1$ choice is equivalent to $c=0$, with $\mu\rightarrow -\mu$.  It follows from (\ref{eom2}) that we have
\bea
f &=& H_1 H_2 + g^2 r^2 \left(H_1^{1+\mu} H_2^{1-\mu}\right)^{\fft{D-2}{D-3}}\cr
&&-\alpha\, r^2 H_2 (H_1-H_2)^{\fft{D-1}{D-3}}\, {}_2F_1[1, \ft{D-2}{D-3}(1+\mu); \ft{2(D-2)}{D-3}; 1 - \ft{H_2}{H_1}]\,.
\label{fsol1}
\eea
The scalar potential can then be derived from (\ref{eom3}), given by
\bea
V &=& - \ft12 (D-2) g^2 e^{\fft{\mu-1}{\nu}\Phi}\Big[
(\mu-1)((D-2)\mu-1) e^{\fft{2}{\nu}\Phi} -2(D-2)(\mu^2-1) e^{\fft{1}{\nu}\Phi}\cr
&&\qquad\qquad\qquad\qquad\quad + (\mu+1)((D-2)\mu+1)\Big]\cr
&&-\ft{(D-3)^2}{2(3D-7)}(\mu+1) \alpha\, e^{-\fft{1}{\nu} (4+\fft{\mu+1}{D-3})\Phi} (e^{\fft{1}{\nu}\Phi} -1)^{3+\fft{2}{D-3}}\cr
&&\times\Big[(3D-7) e^{\fft{1}{\nu}\Phi}\,{}_2F_1[2,1+\ft{(D-2)(\mu+1)}{D-3};
3+\ft{2}{D-2};1 - e^{\fft1{\nu}\Phi}] +\cr
&&\quad - \big((3D-7) + (D-2)(\mu-1)\big)\,
{}_2F_1[3,2+\ft{(D-2)(\mu+1)}{D-3};4+\ft{2}{D-2};1 - e^{\fft1{\nu}\Phi}]
\Big]\,,\label{genpot}
\eea
where
\be
\Phi=\sqrt{\ft{2(D-3)}{D-2}}\,\phi\,.\label{Phidef}
\ee
It is important to note that the parameters $(q_1,q_2)$ disappear in the scalar potential, and hence they are integration constants from solving the equations of motion.  On the other hand, the constants $\mu^2+\nu^2=1$, $g^2$ and $\alpha$ in the function $f$ also appear in the scalar potential.  These parameters specify a class of theories and they do not participate in the first law of thermodynamics.

\subsection{Local solutions in five and four dimensions}

In five and four dimensions, the hypergeometric functions in the scalar potential and in the metric function $f$ become simple. In $D=5$, we have
\bea
V&=& -\ft32(g^2 -\alpha ) e^{\fft{-2(1-\mu)}{\sqrt3\,\nu}\phi}
\Big((\mu-1)(3\mu-1) e^{\fft4{\sqrt3\,\nu}\phi} -
6(\mu^2-1) e^{\fft2{\sqrt3\,\nu}\phi} + (\mu+1)(3\mu+1)\Big)\cr
&&-6\alpha e^{-\fft{1+\mu}{\sqrt3\,\nu}\phi} \Big((\mu+1) e^{\fft2{\sqrt3\,\nu}\phi} -\mu+1\Big)\,,\cr
f&=& H_1 H_2 + (g^2 - \alpha) r^2 H_1^{\fft32(1+\mu)} H_2^{\fft32(1-\mu)}
+\ft12\alpha r^2 H_1 H_2 \big((3\mu+1)H_1 - (3\mu-1)H_2\big)\,.
\eea
Note that the $\alpha$ above was scaled by a factor of $(9\mu^2-1)/8$ compared to the results in the general dimensions. Thus the above result is applicable for $\mu^2\ne\ft19$.  For special values $\mu=\pm 1/3$, we have
\bea
V&=& -4(g^2 -\alpha\phi)\big(e^{-\fft{2}{\sqrt6}\phi} + 2 e^{\fft1{\sqrt6}\phi}\big)-\ft4{\sqrt6} \alpha e^{-\fft2{\sqrt6}\phi}
\big(e^{\sqrt6\phi} + 4 e^{\fft{\sqrt6}{2}\phi} -5\big)\,,\cr
f&=& H_1 H_2 + \big(g^2\mp\ft2{\sqrt6} \alpha \log(\ft{H_1}{H_2})\big) r^2 H_1^2 H_2 \pm \ft2{\sqrt6} \alpha r^2 H_1 H_2 (H_1-H_2)\,.
\eea
Note that the scalar potential with $\alpha=0$ for $\mu^2=\ft19$ arise also in five-dimensional gauged supergravity. (See {\it e.g.}~\cite{tenauthor}.)

In $D=4$, for generic $\mu$, we have
\bea
V &=& -(g^2 -\alpha) e^{-\fft{1-\mu}{\nu}\phi}\Big((\mu-1)(2\mu-1) e^{\fft{2}{\nu}\phi} - 4 (\mu^2-1) e^{\fft{1}{\nu} \phi} +
(\mu+1)(2\mu+1)\Big)\cr
&& -\alpha e^{-\fft{1+\mu}{\nu}\phi}\Big((\mu+1)(2\mu+1) e^{\fft{2}{\nu}\phi} - 4 (\mu^2-1) e^{\fft{1}{\nu} \phi} +
(\mu-1)(2\mu-1)\Big)\,,\cr
f &=& H_1 H_2 + (g^2 -\alpha) r^2 H_1^{2(1+\mu)} H_2^{2(1-\mu)}\cr
&& + \alpha r^2 H_1 H_2 \Big(\mu(2\mu+1)H_1^2 - (4\mu^2-1) H_1 H_2 + \mu(2\mu-1) H_2^2\Big)\,,
\eea
where have scaled $\alpha$ by a factor of $\mu(4\mu^2-1)/3$.  For special values of $\mu=0,\pm\ft12$, we have
\bea
\mu=0:&&\quad V=-2(g^2 - \alpha\phi) (2 +\cosh\phi) - 6\alpha \sinh(\phi)\,,\cr
&&\quad f=H_1H_2+ (g^2 -\alpha\log\ft{H_1}{H_2})r^2 H_1^2 H_2^2 + \ft12 \alpha H_1 H_2 (H_1^2 - H_2^2)\,,\cr
\mu=\pm\ft12:&&\quad V=-6(g^2 -\alpha \phi)\cosh\ft{\phi}{\sqrt3} -
\ft{\sqrt3}2\alpha (9\sinh\ft{\phi}{\sqrt3} + \sinh\sqrt3\phi)\,,\cr
&&\quad f= H_1 H_2 + (g^2 -\ft{\sqrt3}{2}\alpha \log\ft{H_1}{H_2}) r^2 H_1^3 H_2\cr
&&\quad\qquad\qquad + \ft{\sqrt3}{4}\alpha r^2 (H_1-H_2)(3H_1-H_2) H_1 H_2\,.
\eea
To be precise the last $f$ was given for the case $\mu=\ft12$. For $\mu=-\ft12$, we need interchange $H_1$ with $H_2$. Note that the scalar potentials with $\alpha=0$ for $\mu=0,\pm \ft12$ arise also in four-dimensional gauged supergravity. (See {\it e.g.}~\cite{tenauthor}.)

The four-dimensional scalar potential and the local solutions were also obtained in \cite{Anabalon:2013sra}, but in a different coordinate system $x$ specified in (\ref{rtox}).

\subsection{The property of the scalar potential}

The scalar potential we obtained in (\ref{genpot}) has a stationary point
at $\phi=0$, with
\be
V_0\equiv V(\phi=0) = -(D-1)(D-2) g^2\,.\label{cosmoconst}
\ee
The effective cosmological constant is independent of the parameter $\alpha$.  We made a choice that the cosmological constant is negative, so that the solution is asymptotically AdS.  We can also set it zero ($g=0$), or positive ($g^2<0$), giving rise to solutions that is asymptotic flat or de Sitter (dS) respectively.  Performing the Taylor expansion of the potential around $\phi=0$, we find
\be
V=-(D-1)(D-2) g^2 - (D-3) g^2 \phi^2 + \fft{\sqrt2(D-3)^{\fft32}(D-4)\mu g^2}{3\sqrt{D-2}\,\nu} \phi^3 + \cdots
\ee
The $\alpha$ contribution comes at the order of $\phi^{3+\fft{2}{D-3}}$.  Thus the linearized equation for the dilaton around the AdS vacuum is given by (\ref{kgeom}).
This implies that the scalar $\phi$ is massless when $g=0$, and also massless in the gauged supergravity sense when $g\ne 0$, with the asymptotic falloff $1/r^{D-3}$.  This was in fact assumed in our original ansatz.

The scalar potential with $\alpha=0$ was already obtained in \cite{Lu:2013eoa}.  To see this, let us define the notations used in \cite{Lu:2013eoa}:
\be
a_i^2=\fft{4}{N_i} - \fft{2(D-3)}{D-2}\,,\qquad i=1,2\,,\label{aidef}
\ee
where
\be
N_1=\fft{D-2}{D-3} (1+\mu)\,,\qquad N_2=\fft{D-2}{D-3} (1-\mu)\,.\label{Ni}
\ee
To be specific
\be
a_1=-\sqrt{\ft{2(D-3)}{D-2}} \fft{\mu-1}{\nu}\,,\qquad
a_2=-\sqrt{\ft{2(D-3)}{D-2}} \fft{\mu+1}{\nu}\,.
\ee
Note that $a_1 a_2 = -2(D-3)/(D-2)$.  It is then straightforward to see that the scalar potential can be cast into the form presented in \cite{Lu:2013eoa}.  It was shown in \cite{Lu:2013eoa} that the scalar potential can be expressed in terms of a super-potential.  In the present notation, we have
\bea
V &=& \left(\fft{dW}{d\phi}\right)^2 - \fft{D-1}{2(D-2)} W^2\,,\cr
W &=& \ft{1}{\sqrt2}(D-2)\Big( (\mu+1) e^{\fft{\mu-1}{\nu}\Phi} -
(\mu-1) e^{\fft{\mu+1}{\nu}\Phi}\Big)\,,
\eea
where $\Phi$ is defined in (\ref{Phidef}).  As was cataloged in \cite{Lu:2013eoa}, the scalar potentials with integer $N_1$ can be embedded in various gauged supergravities in low-lying dimensions.  Interestingly in four dimensions, some single-scalar truncations of $\omega$-deformed ${\cal N}=8$ gauged supergravity can also arise when $\alpha\ne 0$ \cite{Anabalon:2013eaa}.

\subsection{Global analysis and black hole thermodynamics}

We now study the global structure of our solutions.  First we examine the asymptotic behavior and determine the mass.  The mass is independent of whether the metric has a horizon or a naked singularity.  Let us write the metric in the following Schwarzschild coordinates
\be
ds^2 = - h dt^2 + \fft{d\rho^2}{\tilde h} + \rho^2 d\Omega^2_{D-3}\,.
\label{standardmetric}
\ee
For large $\rho$, we find
\bea
&&\phi = \sqrt{\ft{D-2}{2(D-3)}}\,\nu \Big(\fft{q}{\rho^{D-3}} + \fft{\mu q^2}{2\rho^{2(D-3)}} + \cdots\Big)\,,\qquad
h = g^2 \rho^2 + 1 - \fft{\mu q + \alpha q^\fft{D-1}{D-3}}{\rho^{D-3}} + \cdots\,,\cr
&&\tilde h-h = \fft{g^2(1-\mu^2)q^2}{4\rho^{D-3}} + \cdots\,.
\label{asymptotic}
\eea
where $q=q_1-q_2$.  Thus the mass of the solution can be read off and given by
\be
M=\ft{D-2}{16\pi}q (\mu + \alpha q^{\fft{2}{D-3}})\omega_{D-2}\,.\label{mass}
\ee
where $\omega_{D-2}$ is the volume of the unit round $S^{D-2}$. Note that $q_1$ and $q_2$ enters the metric (\ref{standardmetric}) only through the combination of $q$.  This is expected and as we remarked earlier, the solution only has one non-trivial parameter.  The Schwarzschild-Tangherlini AdS black hole can be obtained by taking the following limit:
\be
q_1\rightarrow \ft12 \epsilon\,,\qquad q_2\rightarrow -\ft12 \epsilon\,,\qquad \alpha \rightarrow m \epsilon^{-\fft{D-1}{D-3}}\,,\qquad
\epsilon\rightarrow 0\,.
\ee
It is clear that in this limit, we have $\phi=0$, $H=1$ and
\be
f=g^2 r^2 + 1 - \fft{m}{r^{D-3}}\,.\label{adsbh}
\ee
Without loss of generality, we shall set $q_2=0$ in the remaining discussion of this section.

The general solution has a curvature power-law singularity at $r=0$. (Note that we have set $q_2=0$ now. For $q_2\ne 0$, the curvature singularity is at $r=-{\rm min}\{q_1,q_2\}$.)  The singularity will be shielded by a horizon if the function $f$ has a positive root.  If we consider $\alpha=0$, then the function will be positive definite from $r=0$ to $r=\infty$.  Thus for the solution to describe a black hole, we must have non-vanishing $\alpha$.  A sufficient condition for a black hole is that $f$ becomes negative as $r$ approaches 0.  Then there must be a positive root since $f$ is positive at asymptotic infinity.  As $r\rightarrow 0$, we find
\be
f=\fft{q_1}{r^{D-3}} \Big(1 - \fft{(D-1)\alpha q_1^{\fft{2}{D-3}}}{1-(D-2)\mu}\Big) + \cdots\,.
\ee
Thus a sufficient condition for the solution to describe a scalar black hole is
\be
1 - \fft{(D-1)\alpha q_1^{\fft{2}{D-3}}}{1-(D-2)\mu}<0\,.\label{bhcons}
\ee
In deriving the result, we made use of the following hypergeometric function identity
\be
{}_2F_1[1,\ft{(D-2)(1+\mu)}{D-3}; \ft{2(D-2)}{D-3};1]=
\ft{D-1}{1-(D-2)\mu}\,.
\ee
Since $q_1>0$ is an integration constant, it follows that for $\alpha>0$ and
$1-(D-2)\mu>0$, one can always find appropriate $q_1$ so that the
condition (\ref{bhcons}) is satisfied. We find that in this case the condition (\ref{bhcons}) implies that the mass is positive:
\be
M> \fft{(D-2)\Omega_{D-2}}{16\pi(D-1)} (1+\mu)q_1 > 0\,.
\ee
We shall not go through the tedious classification of the parameter space of all black hole solutions. It suffices for us to use a few numerical numbers satisfying the condition (\ref{bhcons}) to establish that black holes do exist. We are now in the position to derive the first law of thermodynamics.  Let $r_0$ denote the largest positive root of $f(r)$, the temperature and the entropy
can be calculated straightforwardly using the standard technique:
\be
T=\fft{f'(r_0)}{4\pi \big(H(r_0)\big)^{\fft{D-2}{2(D-3)}}}\,,\qquad
S=\ft14 r_0^{D-2} \big(H(r_0)\big)^{\fft{D-2}{2(D-3)}} \omega_{D-2}\,.
\ee
The first law of thermodynamics
\be
dM=T dS\label{firstlaw}
\ee
can be easily verified since it turns out that the hypergeometric function is not involved.  In fact, we have a simple expression for $f'(r_0)$, namely
\be
f'(r_0)=r_0^{-(D-1)}\Big(q_1\big((D-2)\mu-1 + \alpha(D-1)q_1^{\fft{2}{D-3}}\big)
-2 r_0^D\Big)\,.
\ee
The free energy is given by
\be
F=M - T S = \fft{\omega_{D-2}}{16\pi}\Big(q_1 + 2r_0^{D-3} - \alpha
q_1^{\fft{D-1}{D-3}}\Big)\,.
\ee
The minus sign implies that a global phase transition could occur when
the free energy change the sign. If we can adjust the parameters such
that $f'(r_0)$ vanishes, the solution becomes an extremal black hole with the AdS$_2\times S^{D-2}$ near-horizon geometry.  We find that the condition $f(r_0)=0=f'(r_0)$ can only be satisfied with negative $g^2$, {\it i.e.} the positive cosmological constant.  Thus hairy extremal black holes with zero temperature exist with asymptotic dS spacetimes in our theories.

    To conclude, for suitable choice of the parameter $\alpha$, our
solutions describe scalar hairy black holes. We verify the first law of thermodynamics.  For positive $g^2$, corresponding to asymptotic AdS, there is only one real positive root for $f$ and hence there is no extremal limit. For negative $g^2$, corresponding to asymptotic dS, there can be two real positive roots for $f$ and extremal black holes can exist when the two roots coalesce.

It is important to emphasize that our black hole solution has only one non-trivial parameter, namely the mass.  In a given theory with some appropriate $\alpha$ and $\mu$, there can exist two black holes with the same mass.  One is the usual (A)dS Schwarzschild black hole (\ref{adsbh}) and other is our scalar black hole.  They have different temperature and entropy, but satisfy the same first law of thermodynamics (\ref{firstlaw}). The uniqueness is thus broken and our black holes are hairy.

\subsection{Black holes with different topologies}

For asymptotic AdS (or dS) solutions, one can replace the foliating sphere with the metric $d\Omega_{D-2}^2$ with those of a torus and hyperbolic space.  Let us denote $d\Omega_{D-2,k}^2$ as the metric of a unit sphere, torus and hyperbolic space corresponding to $k=1,0,-1$ respectively.  Following the scaling procedure described in \cite{Lu:2013eoa}, we obtain black hole solutions with different topologies.  The only change to the metric function is that $f$ is now given by
\bea
f &=& k H_1 H_2 + g^2 r^2 \left(H_1^{1+\mu} H_2^{1-\mu}\right)^{\fft{D-2}{D-3}}\cr
&&-\alpha\, r^2 H_2 (H_1-H_2)^{\fft{D-1}{D-3}}\, {}_2F_1[1, \ft{D-2}{D-3}(1+\mu); \ft{2(D-2)}{D-3}; 1 - \ft{H_2}{H_1}]
\eea

\section{Charged black holes}

Having obtained the scalar hairy black holes, we now add electric charges to these solutions.  A natural class of theories to consider is
\be
e^{-1} {\cal L} = R - \ft12(\partial\phi)^2 - \ft14 e^{b_1\phi} (F^1)^2 -
\ft14 e^{b_2\phi} (F^2)^2 - V(\phi)\,.
\ee
where $F^1=dA^1$ and $F^2=dA^2$.  The motivation of introducing two vector fields is two fold.  One is that it can be easily reduced to Einstein-Maxwell theory provided that $b_1$ and $b_2$ have the opposite sign.  The other is that in four dimensions, when $b_1=-b_2$, the two electric charges associated with $F^i$ can be viewed as the electric and magnetic charges of one single field strength.  Note that when we constructed the scalar black holes in the previous section, we introduced functions $H_1$ and $H_2$, which over parameterized the solutions.  As we shall see, the results have the advantage to incorporate the two charges of $F^i$ in a symmetric fashion.

The electrically charged spherically-symmetric ansatz is given by
\bea
ds^2 &=& -H^{-1} f dt^2 + H^{\fft{1}{D-3}} \Big(\fft{dr^2}{f} + r^2 d\Omega_{D-2}^2\Big)\,,\cr
F^i & =& \fft{\lambda_i e^{-b_i\phi}}{r^{D-2} H} dt\wedge dr\,.
\eea
It turns out that the equations of motion (\ref{eom1}) and (\ref{eom3}) are not modified at all.  This implies that we can use the same relation of $\phi$ and $H$ established previously.  The equation (\ref{eom2}) becoms
\bea
&&\fft{H''}{H}+\fft{H'}{H} \Big(\fft{f'}{f}-\fft{H'}{H} +
\fft{D-2}{r}\Big)+\fft{(D-3)(\lambda_1^2 e^{-b_1\phi} + \lambda_2^2 e^{-b_2\phi})}{(D-2)r^{2(D-2)} H f}\cr
&&-\fft{(D-3)}{(D-2)r^2f}\Big(r^2f'' + (D-4) r f' - 2(D-3)(f-1)\Big)=0\,.
\eea
The charges contribute some new source to the second-order linear differential equation of $f$ and hence $f$ will be modified by these terms.  From the equation (\ref{eom3}) that solves for $V$, which depends only on $f$ and $H$, we conclude that for generic constant parameters $(b_1,b_2)$, the potential $V$ will depend explicitly on $\lambda_i$, even if one can solve for $f$.  Since we are only interested in theories where the charges arise as integration constant, rather than pre-fixed by the theory, we thus require that the extra $f_{\rm ext}$ from the $\lambda_i$ contribution do not modify the scalar potential.  It follows from (\ref{eom3}) that we must have
\be
f_{\rm ext}=\fft{c_1}{r^{D-3}} + \fft{c_2}{r^{2(D-3)}}=\fft{\tilde c_1 H_1 + \tilde c_2 H_2}{r^{D-3}}\,,
\ee
where $c_i$ (or $\tilde c_i$) are integration constants. This restricts the possible dilaton coupling constants $b_1$ and $b_2$,
which turn out to be $b_i=a_i$, where $a_i$'s were defined in (\ref{aidef}). We then find that the charged solutions are given by
\be
f=f_0 + \fft{(1+\mu)\lambda_2^2 H_1 -(1-\mu) \lambda_1^2 H_2}{(D-2)(D-3)(1-\mu^2)(H_1-H_2)r^{2(D-3)}}\,.
\ee
where $f_0$ denotes the $f$ of the pure scalar solution, given in (\ref{fsol1}).  The scalar potential is the same as (\ref{genpot}).  It is straightforward to obtain solutions in other topologies with appropriate insertion of the discrete parameter $k$, as discussed in the end of the last section.

The solutions contain three non-trivial parameters, corresponding to the mass and
two electric charges, given by
\be
M = \ft{(D-2)\omega_{D-2}}{16\pi}\Big\{\mu q + \alpha q^{\fft{D-1}{D-3}} +
\fft{\lambda_1^2(1+\mu) - \lambda_2^2 (1-\mu)}{(D-2)(D-3)(1-\mu^2)q}\Big\}\,,\qquad
Q_i = \ft{\lambda_i}{16\pi} \omega_{D-2}\,,\label{para1}
\ee
where $q=q_1-q_2$.  To make contact with the STU-type two-charge black holes constructed in \cite{Lu:2013eoa}, we define
\be
q_i=m s_i^2\,,\qquad \lambda_i=(D-3)\sqrt{N_i}\, m c_i s_i\,,\label{qilambdai}
\ee
where $c_i=\cosh\delta_i$ and $s_i=\sinh\delta_i$ and $N_i$ are given in (\ref{Ni}).  We then have $H_i=1+m s_i^2/r^{D-3}$ and
\be
f=1 - \fft{m}{r^{D-3}} + g^2r^2 H_1^{N_1} H_2^{N_2} +\alpha\, r^2 H_2 (H_1-H_2)^{\fft{D-1}{D-3}}\, {}_2F_1[1, \ft{D-2}{D-3}(1+\mu); \ft{2(D-2)}{D-3}; 1 - \ft{H_2}{H_1}]\,.
\ee
The mass and charges are now given by
\bea
M &=& \ft{(D-2)\omega_{D-2}}{16\pi}\Big\{m\big(1 + \ft{D-3}{D-2}(N_1 s_1^2 +
N_2 s_2^2)\big) + \alpha \big(m(s_1^2-s_2^2)\big)^{\fft{D-1}{D-3}}\Big\}\,,\cr
Q_i &=& \ft{(D-3)\omega_{D-2}}{16\pi} m \sqrt{N_i}\,c_i s_i\,.\label{para2}
\eea
The solutions with $\alpha=0$ in this parametrization become precisely those in \cite{Lu:2013eoa}.

The two parametrization schemes are not equivalent.  In the parametrization
(\ref{para1}), we can turn off the charges $Q_i\sim \lambda_i$, without turning off
the parameter $q$, the resulting solution is the scalar hairy black hole discussed in the previous section. In the parametrization (\ref{para2}), turning off the charges implies setting $\delta_i=0$, which has the effect of setting $q_i=0$.  The resulting solution is then the Schwarzschild black hole with a mass parameter $m$.  Thus the charged black holes become hairy, involving two branches of solutions for given mass and charges.

\section{Spherical M-branes and D3-branes}

As we mentioned in section 3, when the $\alpha=0$, the neutral scalar solutions no longer possess a horizon and they can be interpreted as domain walls.  In fact, the scalar potential with $\alpha=0$ can be embedded in gauged supergravities in low-lying dimensions for integer $N_1$.  The planar domain walls with $k=0$ are among those constructed in \cite{Cvetic:1999xx}.  The $k=1$ solutions may be interpreted as spherical domain walls.

The supergravity theories in four and seven-dimensions can be obtained from the Kaluza-Klein reduction of eleven-dimensional supergravity on seven or four spheres.  The five dimensional theory can be obtained from five-sphere reduction of type IIB supergravity.  The reduction ansatz for our subset scalar fields of gauged supergravities can be found in \cite{tenauthor}.  Let us consider the simplest case with $q_2=0$ and $N_1=1$.  The metric in $D=11$ and $D=10$ can be easily obtained using the reduction ansatz in \cite{tenauthor}.  We find
\bea
ds_{10}^2 &=&
H^{\fft12}\Big(\fft{dr^2}{f} + r^2(\Delta d\theta^2 + H_1 \sin\theta^2 d\phi^2 + \cos^2\theta d\Omega_3^2)\Big)\cr
&&+H^{-\fft12} \Big(-f dt^2 + d\Omega_{3,k}^2\Big)\,,\qquad H=\fft{1}{g^4r^4\Delta}\,,\qquad f=g^2 + \fft{k}{r^2}\,,\cr
ds_{11}^2 &=& H^{\fft13}\Big(\fft{dr^2}{f} + r^2(\Delta d\theta^2 + H_1 \sin\theta^2 d\phi^2 + \cos^2\theta d\Omega_5^2)\Big)\cr
&&+H^{-\fft23} \Big(-f dt^2 + d\Omega_{2,k}^2\Big)\,,\qquad
H=\fft{1}{g^6r^6 \Delta}\,,\qquad f=g^2 + \fft{k}{r^4}\,,\cr
ds_{11}^2 &=&H^{\fft23}\Big(\fft{dr^2}{f} + r^2(\Delta d\theta^2 + H_1 \sin\theta^2 d\phi^2 + \cos^2\theta d\Omega_2^2)\Big)\cr
&&+H^{-\fft13} \Big(-f dt^2 + d\Omega_{5,k}^2\Big)\,,\qquad
H=\fft{1}{g^3r^3\Delta}\,,\qquad f=g^2 + \fft{k}{r}\,,
\eea
where $\Delta = \sin^2\theta + H_1 \cos^2\theta$ and $H_1=1 + q_1/r^2$ are universal for the three solutions. The solutions with $k=0$ can be viewed distributed D3-brane or M-branes on a disk \cite{Cvetic:1999xx}.  For $k=1$, the solutions may be similiarly interpreted as the spherical D3-brane or M-branes.

\section{Black holes with scalar charges?}

One motivation of our work is to better understand scalar charges in asymptotic AdS geometries. It was shown in \cite{Lu:2013ura} that in four dimensions the first law of thermodynamics can be modified with the full differential $dM$ being shifted by a 1-form $Z\equiv XdY$:
\be
dM \rightarrow dM + Z\,,\qquad Z=\ft1{12} g^2 (2\phi_2 d\phi_1 - \phi_1 d\phi_2)\,.\label{scalarcharge}
\ee
Here $\phi_i$ are the two leading order expansion parameters in the large $\rho$ expansion
\be
\phi = \fft{\phi_1}{\rho} + \fft{\phi_2}{\rho^2} + \cdots
\ee
where $\rho$ is defined in (\ref{standardmetric}).  The result was generalized to general dimensions \cite{Liu:2013gja}. However, there is not yet an independent method of deriving the scalar charge $Y$ and its thermodynamical potential $X$ that would give the 1-form $Z$. The only known examples so far that have such a scalar charge are the Kaluza-Klein AdS dyonic black hole \cite{Lu:2013ura} and its multi-charge generalizations \cite{Chow:2013gba}.

In all our scalar hairy black holes, the first law of thermodynamics (\ref{firstlaw}) is satisfied, without any modification from the scalar charge. It is clear from the asymptotic behavior (\ref{asymptotic}) that for our choice of $\phi$, the quantity $Z$ vanishes identically. The same is true for the ansatz used in \cite{Anabalon:2012ta,Anabalon:2013qua,Anabalon:2013sra,Gonzalez:2013aca}.
It would be interesting to find solutions with non-vanishing $Z$.  Recently a new type of black holes involving solving a set of $SL(n,\mathbb{R})$ Toda equations were constructed in \cite{Lu:2013uia}, which includes the dyonic black hole with no scalar potential as a special case. Inspired by the single-scalar $SL(3,R)$ example, we consider
\bea
&&\phi=\sqrt{\ft{3(D-2)}{8(D-3)}} \log \fft{H_1}{H_2}\,,\qquad
H=\sqrt{H_1H_2}\,,\cr
&&H_i = 1 + \ft{q_i}{r^{D-3}} + \ft{p_i}{r^{2(D-3)}}\,,\qquad
i=1,2.
\eea
The equation (\ref{eom1}) is satisfied provided that
\be
p_1 + p_2=\ft12 q_1 q_2\,.\label{pqrelation}
\ee
The condition was solved with three parameters $(m,\beta_1,\beta_2)$ in
\cite{Lu:2013uia}, and in \cite{Lu:2013ura} for $D=4$.  In $D=4$, the solution can be embedded in gauged supergravity, and it becomes the dyonic AdS black hole \cite{Lu:2013ura}.  We opt to solve the condition (\ref{pqrelation}) by
\be
p_1 = \ft14 q_1q_2 + p_0^2\,,\qquad
p_2 = \ft14 q_1q_2 - p_0^2\,.
\ee
To proceed, let us restrict our attention  in four dimensions.  The $\phi_1$ and $\phi_2$ in the large $\rho$ expansion is now given by
\be
\phi_1 = \ft{\sqrt3}{2} (q_1-q_2)\,,\qquad \phi_2=\ft14(p_0^2 - q_1^2 + q_2^2)\,.
\ee
Thus if $(q_1,q_2,p_0)$ are independent parameters, the 1-form $Z$ in (\ref{scalarcharge}) will be non-vanishing.  For general parameters $(p_0,q_1,q_2)$, we can solve for $f$ using (\ref{eom2}) and we find that result involves logarithmic and $\arctan$ functions:
%\bea
%f&=& g^2r^2 H_1H_2  - \fft{r^2 H_1 H_2 }{p_0^2(4p_0^2-q_1^2)}\Big
%((2p_0-q_1) \log(r-p_0) + (2p_0+q_1) \log(r+p_0)\cr
%&&- 2p_0 \log(r^2 + r q_1 + p_0^2)\Big) + \alpha r^2 H_1 H_2 \Big(
%2(4p_0^2 - q_1^2) \arctan\fft{\sqrt{4p_0^2-q_1^2}}{2r+q_1}\cr
%&& -4p_0\sqrt{4p_0^2-q_1^2} {\rm arctanh}\fft{p_0}{r} +
%q_1\sqrt{4p_0^2-q_1^2}\log\fft{H_1}{H_2}\Big)\,.
%\eea
\bea
f&=& H_1 H_2 + g^2 r^2 H_1 H_2 +\alpha r^2 H_1 H_2 \log(\fft{H_1}{H_2})\cr
&& +\fft{2 r^2 H_1 H_2}{q_1-q_2} \Big(
\fft{\alpha(4p_0^2 + q_1(q_2-q_1))-2}{\sqrt{4p_1-q_1^2}} \arctan
\sqrt{\ft{4p_1-q_1^2}{2r+q_1}}\cr
&&\qquad\qquad\qquad-\fft{\alpha(4p_0^2-q_2(q_2-q_1))-2}
{\sqrt{4p_2-q_2^2}}\arctan
\sqrt{\ft{4p_2-q_2^2}{2r+q_2}}\Big)\,.
\eea
We can then derive the scalar potential using (\ref{eom3}).  However, we find that all the parameters appear explicitly in the scalar potential, with none of them can be fully absorbed into the scalar field.  This implies that there is no free parameters in our solution, but all pre-fixed by the theory.  Thus, although we obtain solutions that we expect to have scalar charge $Y$ and its potential $X$, the discussion of their contribution to the first law of thermodynamics is moot because they are purely constants.

We may also generalize the Kaluza-Klein dyonic AdS black hole
constructed in \cite{Lu:2013ura}.  The solution is complicated and we shall
not repeat it here, but simply adopt the exact same notation
of \cite{Lu:2013ura}.  We generalize the solution (2.1) of \cite{Lu:2013ura}
by letting
\be
f\rightarrow f + f_{\rm ext}\,,\qquad V\rightarrow V + V_{\rm ext}\,.
\ee
we find that the most general non-trivial solution for $f$ is given by
\bea
f_{\rm ext} &=& \alpha \log\fft{H_1}{H_2} -
\fft{2\alpha}{(1-\beta_1\beta_2)\sqrt{\beta_1-\beta_2}}
\Big(\gamma_2\sqrt{\beta_1} \arctan
\fft{\sqrt{\beta_1(\beta_2-\beta_1)}(1-f_0)}
{1-\beta_1-\beta_1(1-\beta_2)f_0}\cr
&&\qquad\qquad+\gamma_1\sqrt{\beta_2} {\rm arctanh}
\fft{\sqrt{\beta_2(\beta_2-\beta_1)}(1-f_0)}
{1-\beta_2-\beta_2(1-\beta_1)f_0}\Big)\,.
\eea
where $H_i$ and $f_0$ are given in \cite{Lu:2013ura}.  The solution
in \cite{Lu:2013ura}, corresponding to $\alpha=0$,
contains three independent parameters $(\mu,\beta_1,\beta_2)$ parameterizing
the mass, electric and magnetic charges.  (There should be no confusion
between the notation of our $\mu$ and the mass parameter $\mu$ in
\cite{Lu:2013ura}.) The scalar potential $V_{\rm ext}$ is
somewhat complicated and we shall not present it here. It depends on the dimensionless parameters $(\beta_1,\beta_2)$, but is
independent of the parameter $\mu$ that has the dimension of mass.  Thus the generalized dyonic AdS black hole
does contain free parameter and the first law of thermodynamics can be analyzed.
Unfortunately, with the dimensionless $\beta_i$ parameters now being fixed,
$\phi_2/\phi_1^2$ must be a pure constant on the ground of dimensional analysis
and hence the 1-form $Z$ is again vanishing. So far the only examples with
non-vanishing $Z$ remain the Kaluza-Klein dyonic AdS black holes \cite{Lu:2013ura,Chow:2013gba}.

\section{Conclusions}

In this paper, we generalized to higher dimensions the construction of scalar black holes that were known by and large in four dimensions.  In this construction, one assumes a specific function for the scalar and then derive the full set of metric functions and also the scalar potential that is responsible for the solution.  We made the coordinate choice that was inspired by the standard ansatz in the $p$-brane construction.  This has the advantage that the asymptotic infinity can be easily seen and also that we can use directly the knowledge in the $p$-brane construction.

We obtained a class of scalar potentials (\ref{genpot}) for a massless scalar $\phi$ in general dimensions, with the coupling parameters $g$ and $\alpha$.  All the potentials have a stationary point at $\phi=0$ that gives rise to a cosmological constant (\ref{cosmoconst}) parameterized by $g$.  We obtained a class of static solutions that describe hairy black holes for some suitable choice of parameters.  Our theory admits also the Schwarzschild-Tangherlini AdS black holes, and for a given mass, they can have different temperature and entropy, but satisfy the same first law of thermodynamics, compared to the new scalar black holes.  Thus the uniqueness is broken.  When $\alpha=0$, the scalar potentials were known previously and they can be expressed in terms of super-potentials.  In low-lying dimensions, some of these potentials can arise from gauged supergravities. The solutions with $\alpha=0$ however develop a naked curvature singularity and become spherical domain walls.  For those can be embedded in gauged supergravities, the domain walls can be lifted and become the spherical M-branes and D3-brane.

     We then introduced two vector fields in the theory, and constructed hairy
charged black holes.  For $\alpha=0$, one branch of the solutions become identical to the charged dilatonic AdS black holes obtained previously, including the ones in STU-like models of gauged supergravities in low-lying dimensions.

     We also made an effort in constructing AdS black holes in
four dimensions that carry non-trivial scalar charges. Unfortunately, all the parameters in one solution are pre-fixed and hence they give no information of the black hole first law of thermodynamics.  In the other solution that generalizes the dyonic AdS black hole of \cite{Lu:2013ura}, although there is one
free integration constant, but it is not enough for the scalar charge
to give the non-vanishing contribution to the first law of thermodynamics.

    Very little is known about scalar hairy black holes and even less is
known about the scalar charges in asymptotic AdS geometries and their implication in the AdS/CFT correspondence.  Further investigation is warranted
in this subject.

\section*{Acknowledgement}

We are grateful to Chris Pope for useful discussions. The research of H.L.~is supported in part by the NSFC grants 11175269 and 11235003.


\begin{thebibliography}{99}

%\cite{Behrndt:1998jd}
\bibitem{Behrndt:1998jd}
  K.~Behrndt, M.~Cveti\v c and W.A.~Sabra,
{\it Nonextreme black holes of five-dimensional ${\cal N}=2$ AdS supergravity,}
  Nucl.\ Phys.\ B {\bf 553}, 317 (1999)
  [hep-th/9810227].
  %%CITATION = HEP-TH/9810227;%%
  %210 citations counted in INSPIRE as of 29 May 2013

%\cite{Duff:1999gh}
\bibitem{Duff:1999gh}
  M.J.~Duff and J.T.~Liu,
{\it Anti-de Sitter black holes in gauged ${\cal N} = 8$ supergravity,}
  Nucl.\ Phys.\ B {\bf 554}, 237 (1999)
  [hep-th/9901149].
  %%CITATION = HEP-TH/9901149;%%
  %126 citations counted in INSPIRE as of 29 May 2013

%\cite{tenauthor}
\bibitem{tenauthor}
  M.~Cveti\v c, M.J.~Duff, P.~Hoxha, J.T.~Liu, H.~L\"u, J.X.~Lu, R.~Martinez-Acosta and C.N.~Pope {\it et al.},
{\it Embedding AdS black holes in ten-dimensions and eleven-dimensions,}
  Nucl.\ Phys.\ B {\bf 558}, 96 (1999)
  [hep-th/9903214].
  %%CITATION = HEP-TH/9903214;%%
  %258 citations counted in INSPIRE as of 28 May 2013

%\cite{Cvetic:1999un}
\bibitem{Cvetic:1999un}
  M.~Cveti\v c, H.~L\"u and C.N.~Pope,
{\it Gauged six-dimensional supergravity from massive type IIA,}
  Phys.\ Rev.\ Lett.\  {\bf 83}, 5226 (1999)
  [hep-th/9906221].
  %%CITATION = HEP-TH/9906221;%%
  %75 citations counted in INSPIRE as of 04 Jun 2013

%\cite{Klemm:2012yg}
\bibitem{Klemm:2012yg}
  D.~Klemm and O.~Vaughan,
{\it Nonextremal black holes in gauged supergravity and the real formulation of special geometry,}
  JHEP {\bf 1301}, 053 (2013)
  [arXiv:1207.2679 [hep-th]].
  %%CITATION = ARXIV:1207.2679;%%
  %9 citations counted in INSPIRE as of 18 Aug 2013

%\cite{stumodels}
\bibitem{stumodels}
  M.J.~Duff, J.T.~Liu and J.~Rahmfeld,
{\it Four-dimensional string-string-string triality,}
  Nucl.\ Phys.\ B {\bf 459}, 125 (1996)
  [hep-th/9508094].
  %%CITATION = HEP-TH/9508094;%%
  %205 citations counted in INSPIRE as of 16 Dec 2013

%\cite{Wu:2011zzh}
\bibitem{Wu:2011zzh}
  S.-Q.~Wu,
{\it General rotating charged Kaluza-Klein AdS black holes in higher dimensions,}
  Phys.\ Rev.\ D {\bf 83}, 121502 (2011)
  [arXiv:1108.4157 [hep-th]].
  %%CITATION = ARXIV:1108.4157;%%
  %10 citations counted in INSPIRE as of 29 May 2013

%\cite{Chow:2011fh}
\bibitem{Chow:2011fh}
  D.D.K.~Chow,
{\it Single-rotation two-charge black holes in gauged supergravity,}
  arXiv:1108.5139 [hep-th].
  %%CITATION = ARXIV:1108.5139;%%
  %1 citations counted in INSPIRE as of 18 Jun 2013

%\cite{Liu:2012ed}
\bibitem{Liu:2012ed}
  H.~Liu, H.~L\"u and Z.-L.~Wang,
{\it $f(R)$ theories of supergravities and pseudo supergravities,}
  JHEP {\bf 1204}, 072 (2012)
  [arXiv:1201.2417 [hep-th]].
  %%CITATION = ARXIV:1201.2417;%%
  %2 citations counted in INSPIRE as of 09 Dec 2013

%\cite{Lu:2013eoa}
\bibitem{Lu:2013eoa}
  H.~L\"u,
  {\it Charged dilatonic AdS black holes and magnetic AdS$_{D-2} \times R^{2}$ vacua,}
  JHEP {\bf 1309}, 112 (2013)
  [arXiv:1306.2386 [hep-th]].
  %%CITATION = ARXIV:1306.2386;%%
  %9 citations counted in INSPIRE as of 09 Dec 2013

%\cite{Cvetic:1996vr}
\bibitem{Cvetic:1996vr}
  M.~Cveti\v c and H.H.~Soleng,
{\it Supergravity domain walls,}
  Phys.\ Rept.\  {\bf 282}, 159 (1997)
  [hep-th/9604090].
  %%CITATION = HEP-TH/9604090;%%
  %236 citations counted in INSPIRE as of 09 Dec 2013

%\cite{Chamblin:1999ya}
\bibitem{Chamblin:1999ya}
  H.A.~Chamblin and H.S.~Reall,
{\it Dynamic dilatonic domain walls,}
  Nucl.\ Phys.\ B {\bf 562}, 133 (1999)
  [hep-th/9903225].
  %%CITATION = HEP-TH/9903225;%%
  %305 citations counted in INSPIRE as of 09 Dec 2013

%%\cite{Kraus:1998hv}
\bibitem{Kraus:1998hv}
  P.~Kraus, F.~Larsen and S.P.~Trivedi,
{\it The Coulomb branch of gauge theory from rotating branes,}
  JHEP {\bf 9903}, 003 (1999)
  [hep-th/9811120].
  %%CITATION = HEP-TH/9811120;%%
  %170 citations counted in INSPIRE as of 09 Dec 2013

%\cite{Freedman:1999gk}
\bibitem{Freedman:1999gk}
  D.Z.~Freedman, S.S.~Gubser, K.~Pilch and N.P.~Warner,
{\it Continuous distributions of D3-branes and gauged supergravity,}
  JHEP {\bf 0007}, 038 (2000)
  [hep-th/9906194].
  %%CITATION = HEP-TH/9906194;%%
  %216 citations counted in INSPIRE as of 18 Dec 2013

%\cite{Cvetic:1999xx}
\bibitem{Cvetic:1999xx}
  M.~Cveti\v c, S.S.~Gubser, H.~L\"u and C.N.~Pope,
  {\it Symmetric potentials of gauged supergravities in diverse dimensions and Coulomb branch of gauge theories,}
  Phys.\ Rev.\ D {\bf 62}, 086003 (2000)
  [hep-th/9909121].
  %%CITATION = HEP-TH/9909121;%%
  %79 citations counted in INSPIRE as of 09 Dec 2013

%\cite{Anabalon:2012ta}
\bibitem{Anabalon:2012ta}
  A.~Anabalon,
{\it Exact black holes and universality in the backreaction of non-linear sigma models with a potential in (A)dS$_4$,}
  JHEP {\bf 1206}, 127 (2012)
  [arXiv:1204.2720 [hep-th]].
  %%CITATION = ARXIV:1204.2720;%%
  %22 citations counted in INSPIRE as of 09 Dec 2013

%\cite{Anabalon:2013qua}
\bibitem{Anabalon:2013qua}
  A.~Anabalon, D.~Astefanesei and R.~Mann,
{\it Exact asymptotically flat charged hairy black holes with a dilaton potential,}
  JHEP {\bf 1310}, 184 (2013)
  [arXiv:1308.1693 [hep-th]].
  %%CITATION = ARXIV:1308.1693;%%
  %5 citations counted in INSPIRE as of 09 Dec 2013

%\cite{Anabalon:2013sra}
\bibitem{Anabalon:2013sra}
  A.~Anabalon and D.~Astefanesei,
{\it On attractor mechanism of $AdS_{4}$ black holes,}
  Phys.\  Lett.\  B {\bf 727}, 568 (2013)
  [arXiv:1309.5863 [hep-th]].
  %%CITATION = ARXIV:1309.5863;%%
  %4 citations counted in INSPIRE as of 09 Dec 2013

%\cite{Gonzalez:2013aca}
\bibitem{Gonzalez:2013aca}
  P.A.~Gonzalez, E.~Papantonopoulos, J.~Saavedra and Y.~Vasquez,
{\it Four-dimensional asymptotically AdS black holes with scalar hair,}
  arXiv:1309.2161 [gr-qc].
  %%CITATION = ARXIV:1309.2161;%%

%\cite{Acena:2013jya}
\bibitem{Acena:2013jya}
  A.~Acena, A.~Anabalon, D.~Astefanesei and R.~Mann,
{\it Hairy planar black holes in higher dimensions,}
  arXiv:1311.6065 [hep-th].
  %%CITATION = ARXIV:1311.6065;%%
  %1 citations counted in INSPIRE as of 09 Dec 2013

%\cite{Lu:2013ura}
\bibitem{Lu:2013ura}
  H.~L\"u, Y.~Pang and C.N.~Pope,
{\it AdS dyonic black hole and its thermodynamics,}
  JHEP {\bf 1311}, 033 (2013)
  [arXiv:1307.6243 [hep-th]].
  %%CITATION = ARXIV:1307.6243;%%
  %7 citations counted in INSPIRE as of 09 Dec 2013

%\cite{Duff:1996hp}
\bibitem{Duff:1996hp}
  M.J.~Duff, H.~L\"u and C.N.~Pope,
{\it The black branes of M theory,}
  Phys.\ Lett.\ B {\bf 382}, 73 (1996)
  [hep-th/9604052].
  %%CITATION = HEP-TH/9604052;%%
  %126 citations counted in INSPIRE as of 09 Dec 2013

%\cite{Cvetic:1996gq}
\bibitem{Cvetic:1996gq}
  M.~Cveti\v c and A.A.~Tseytlin,
{\it Nonextreme black holes from nonextreme intersecting M-branes,}
  Nucl.\ Phys.\ B {\bf 478}, 181 (1996)
  [hep-th/9606033].
  %%CITATION = HEP-TH/9606033;%%
  %141 citations counted in INSPIRE as of 09 Dec 2013

%\cite{Anabalon:2013eaa}
\bibitem{Anabalon:2013eaa}
  A.~Anabalon and D.~Astefanesei,
{\it Black holes in $\omega$-defomed gauged $N=8$ supergravity,}
  arXiv:1311.7459 [hep-th].
  %%CITATION = ARXIV:1311.7459;%%

%\cite{Chow:2013gba}
\bibitem{Chow:2013gba}
  D.D.K.~Chow and G.~Comp\`ere,
{\it Dyonic AdS black holes in maximal gauged supergravity,}
  arXiv:1311.1204 [hep-th].
  %%CITATION = ARXIV:1311.1204;%%
  %2 citations counted in INSPIRE as of 14 Dec 2013

%\cite{Liu:2013gja}
\bibitem{Liu:2013gja}
  H.-S.~Liu and H.~L\"u,
{\it Scalar Charges in Asymptotic AdS Geometries,}
  arXiv:1401.0010 [hep-th].
  %%CITATION = ARXIV:1401.0010;%%

%\cite{Lu:2013uia}
\bibitem{Lu:2013uia}
  H.~L\"u and W.~Yang,
{\it $SL(n,\mathbb{R})$-Toda black holes,}
  Class.\ Quant.\ Grav.\  {\bf 30}, 235021 (2013)
  [arXiv:1307.2305 [hep-th]].
  %%CITATION = ARXIV:1307.2305;%%
  %2 citations counted in INSPIRE as of 09 Dec 2013

\end{thebibliography}
\end{document}